\documentclass[final,times,twocolumn]{elsarticle}

\usepackage{amssymb,amsmath}
\usepackage{upgreek}

\usepackage{tabularx}% Include figure files\cite{
\usepackage{dcolumn}% Align table columns on decimal point
\usepackage{bm}% bold math
\usepackage{nicefrac}%for in-line fractions

\renewcommand{\figurename}{Fig.}
\newcommand{\figurenames}{Figs.}

\usepackage[usenames,dvipsnames]{color}

\journal{Carbon}
\usepackage{newfloat}
\DeclareFloatingEnvironment[name={Supplementary Figure}]{suppfigure}
\begin{document}

\begin{frontmatter}



\title{Influence of Interfaces on the Transport Properties of  Graphite revealed
by Nanometer Thickness Reduction}


\author{Mahsa Zoraghi$^{a,}$\footnote{Current address: Department
  of Neurophysics, Max Planck Institute for Human Cognitive and Brain
 Sciences, 04103 Leipzig, Germany}}

\author{Jos{\'e} Barzola-Quiquia$^a$}

\author{Markus Stiller$^a$}

\author{Pablo D. Esquinazi$^{a,}$\footnote{Corresponding author, email: esquin@physik.uni-leipzig.de, Tel.: +493419732751}}
\address[label1]{Division of Superconductivity and Magnetism, Felix-Bloch Institute for Solid-state
  Physics, University of Leipzig, 04103 Leipzig, Germany}
  
\author{Irina Estrela-Lopis$^b$} 
\address[label3]{Institute of Medical Physics
  and Biophysics, University of Leipzig, 04107 Leipzig, Germany}

\begin{abstract}
We investigated the influence of  thickness reduction
on the transport properties of graphite microflakes. Using oxygen plasma etching
we decreased the thickness of highly oriented pyrolytic graphite (HOPG) microflakes 
from $\sim 100$~nm to $\sim 20$~nm systematically. Keeping current
and voltage electrodes intact, the electrical resistance $R(T)$, the magnetoresistance (MR)
and Raman spectra were measured in every individual sample and after each etching step of a few nm. 
The results show that $R(T)$
and MR can increase or decrease with the sample thickness in a
non-systematic way. The results 
indicate that HOPG samples are inhomogeneous materials, in agreement with scanning
transmission electron microscopy images and X-ray diffraction data. Our results further
indicate that the quantum oscillations in the MR are not an intrinsic property
 of the ideal graphite structure but
their origin is related to internal conducting interfaces. 

\end{abstract}

%\begin{keyword}

%\PACS 73.40.-c, 75.47.-m,81.05.uf

%\end{keyword}

\end{frontmatter}


\section{Introduction}
\label{introduction}

%\maketitle

Experimental
results of the  temperature
dependence of the electrical resistance of bulk highly oriented graphite
samples of good quality reported in the literature show  usually a metallic-like behavior. This
behavior, together with Shubnikov-de Haas oscillations, were taken
as  evidence for the existence of a three dimensional Fermi surface and
the band structure of graphite was proposed using tight binding calculations
with up to seven free coupling constants \cite{KELLY}.
In the last years, however, new experimental results on
graphite bulk and mesoscopic graphite samples have shown that the temperature
dependence of the  resistance~\cite{oha00,kim05,bar08} and the  resistivity~\cite{bar08,zor17} have
a thickness dependence incompatible with the
assumed
metallic- or semimetalliclike behavior. High resolution X-ray diffraction (XRD) data  and scanning transmission
electron microscopy (STEM) pictures of highly ordered pyrolytic
graphite (HOPG) as well as natural graphite samples revealed a system
far from being homogeneous and single phase \cite{chap7}, as assumed in
most of the common literature in the past.
The existence of two stacking orders, Bernal and in a smaller amount rhombohedral, the interfaces between those orders and the
interfaces between twisted  crystalline regions around a common $c-$axis  \cite{pre16,chap7},
make usual bulk graphite samples an inhomogeneous system, structurally as well as
electronically.

A recently published study on the temperature dependence of the resistance of
more than twenty samples of different origins and thicknesses obtained in four
different laboratories, provided a semiquantitative explanation for a rather
complicated temperature behavior \cite{zor17}.  The simple model assumes three
different resistors in parallel: One from the Bernal and one from the
rhombohedral stacking orders and a third one due to interfaces \cite{zor17}, a model similar to the
one proposed in \cite{gar12} but including the rhombohedral stacking contribution. The samples with
different thicknesses were prepared in those studies  by exfoliation \cite{oha00,kim05,bar08,zor17}. It means that each
sample  after reaching a given thickness, was electrically contacted with four electrodes
to measure the resistance.  The  resistivity  vs. thickness of those samples reveals a
clear tendency, namely, towards an increase by reducing the thickness. The observed increase
in the resistivity is not related to an extra disorder produced during the sample preparation
process,  as Raman measurements indicate \cite{zor17}. For a
thickness below $\sim 100~$nm, the temperature dependence of $R(T)$ tends to be more
semiconducting-like   \cite{bar08,zor17}.

A closer look
at the reported thickness dependence of the resistivity \cite{bar08,zor17} reveals a certain
scattering around the main tendency. This scattering is much larger than the  experimental  
errors and, as we will show in this work, is related
to the inhomogeneity of each graphite sample. This knowledge is of importance if
one wants to compare the behavior of nominally ``similar" samples prepared from the same
or different batches.

\begin{figure}
\includegraphics[width=\columnwidth]{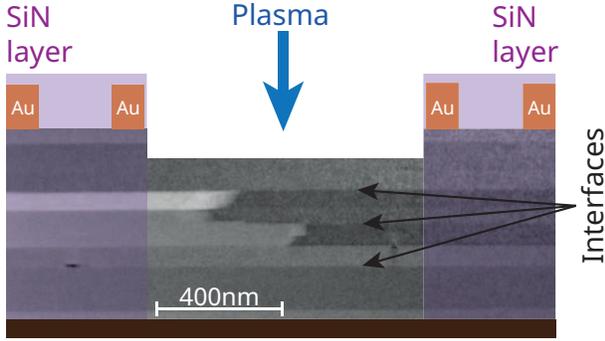}
\caption{\label{tem} Transmission electron microscope pictures of a thin lamella taken from
the same batch as the samples measured in this study. The sketch around the picture shows the
position of the electrodes at the top graphene layer. The usual width and length of the microflakes  were
a few micrometers with thickness below $\sim 100~$nm, see text for more details.}
\end{figure}

To clarify this point further, let us use as an example
a STEM picture of a HOPG sample
with grade A (rocking curve width $\sim 0.4^\circ$), as studied in this work, see
Fig.~\ref{tem}. This STEM picture was taken with the e-beam parallel to the
graphene planes of the sample. The different grey colors indicate either different
stacking and/or twisted crystalline regions around the common $c-$axis. As pointed
out in recent studies \cite{bar08,gar12,zor17} some of those interfaces
show a metallic (and/or superconducting) behavior (see \cite{chap7} for a review).
Therefore, if the current input and voltage electrodes are localized, as usual, at the
top free graphene layer of the sample, due to the large anisotropy in the
resistivity $\rho_c >> \rho_{a,b}$  the near surface region at the top of
 the sample provides the main
contribution to the measured voltage.

In other words,
if an interface due to, e.g., twisted graphene layers (as clearly measured in early studies~\cite{kuw90}),
is located under the top surface layer, 
 the measured voltage will be smaller than in
the case where such an interface is localized deeper in the sample interior.
This is expected because  twisted graphene layers \cite{kuw90,bri12,mil10,San-Jose2013,chap7}
as well as at the interfaces between Bernal and rhombohedral stackings \cite{kopbook,mun13}  show a fundamentally
different and higher density of states than  ideal Bernal stacking. This
is basically the reason for the existence of higher conductivity  most of the relatively large and thick 
graphite samples show \cite{gar12,zor17,chap7}. 

\textcolor{black}{On average and from the STEM pictures obtained in HOPG samples of ZYA grade, the density of interfaces is of the order of $\sim 2 \times 10^{-3}$, which means 
$\sim$ two interfaces every $10^3$ graphene layers. However,  STEM pictures, obtained 
at relatively low electron energies of the order of 30~keV, have two restrictions: On one side they have a finite resolution, which does not always allow us to recognize interfaces between regions twisted with a very small angle. On the other side those STEM pictures scan in general only a very small part of the whole
HOPG sample, roughly $\sim 1~\upmu$m$^2$ area of a region parallel to the $c$-axis. The measured samples, however, are several microns long. Within this length, even for a constant density of interfaces, interfaces exist at different distances from the surface. This fact together with the grain boundaries that exist at the end of the single stacking regions, see, e.g., the STEM picture in Fig.~\ref{tem}, mean that the input current does not always flow at one single and short interface closest to the surface, but it can flow through different interfaces at different locations. Etching a few nm from the surface can or cannot remove most of the interfaces that affect the measured voltage. In other words, the observed changes in the resistance may appear as due to a larger interface density.}

\textcolor{black}{From XRD data \cite{pre16} we know that the percentage ratio between the rhombohedral and Bernal stacking orders is $< 5\%$  to $\sim 20\%$ in bulk samples of the selected ZYA grade sample. From STEM pictures we guess that the layers of the rhombohedral stacking order can have a thickness between a few nm to 20~nm, whereas the Bernal stacking can have a thickness between a few tens of nm to $\gtrsim 400~$nm. As will be clear in the discussion, although our phenomenological model fits extremely accurate the temperature dependence of the resistance $R(T)$, it is not possible to estimate the absolute thickness of the layers from the experimental data but  the ratio between the two stacking orders.}

The main aim of this work was to study the behavior of the resistance
as a function of thickness by thinning the samples in steps of a few nanometers.
In this work we not only studied the changes in the resistivity and
its temperature dependence but also the magnetoresistance.
In contrast to all other published studies we investigated this behavior without changing the
corresponding electrodes. This allowed us to observe a clearly sample dependent  behavior
of the resistance with thickness and, in some cases, even a non monotonous one.
A gentle oxygen plasma etching procedure was used to decrease
systematically the sample thickness between the protected
voltage and current electrodes.  The appearance of disorder at the sample free surface
generated during the etching
process was investigated by means of confocal Raman spectroscopy.
The overall results support the view that the graphite samples are,
electronically speaking, highly inhomogeneous with an important contribution
from internal interfaces.

 \begin{figure}
\includegraphics[width=\columnwidth]{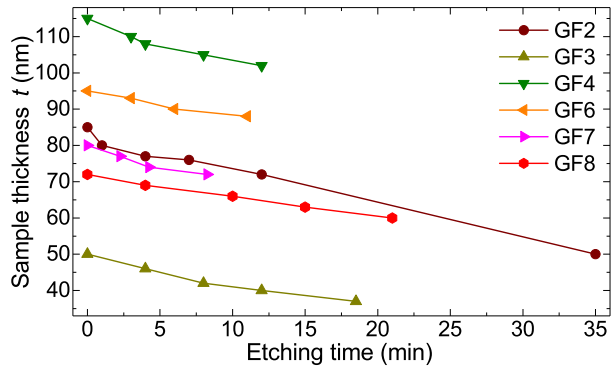}
\caption{\label{figxx} Sample thickness measured by atomic force microscopy (AFM) vs. etching time in minutes
for six different mesoscopic graphite samples.
The etching of the graphite flakes was done using an oxygen plasma chamber, 
power and pressure were kept constant in time and for each etching step.
 (see main text for details).}
\end{figure}

\section{Experimental procedures}

The mesoscopic graphite samples used for our experiments were obtained from a bulk
HOPG  material from Advanced Ceramics, with a rocking curve of $0.4^{\rm o}$
and metallic impurities in the ppm range~\cite{spe14}.
The flakes were produced using a rubbing method already
described in previous publications~\cite{bar08}.  After pre-selecting the flakes with
an optical microscope they were attached on the top of silicon substrates caped
with 150~nm thick insulating silicon nitride (Si$_3$N$_4$). After further selection
of suitable samples, electron beam lithography was used to print the
structures for the electrodes, which were sputtered with a bilayer of
Cr$/$Au with a thickness of $\approx 5$~nm and $\approx 30$~nm,
respectively.  The main contact area of the electrodes was at the top
of the sample. Although the Cr$/$Au deposited film can touch part of the
edges of the flakes, the results presented in this work indicate
that the main part of the current input takes place at the large
top electrode area.

Oxygen plasma etching was employed
to reduce the thickness of the samples.
The etching was done using passive oxygen plasma
etching (ZEPTO), with a gas flow of $\approx80$~sccm. Experiments
were performed at a pressure of $\approx 1.1 \times
10^{-2}$~mbar. The oxygen plasma was generated by a microwave generator at a power
of 50 W and a frequency of 13.56~MHz. Samples were exposed to the
oxygen plasma for different times.  The longest treatment
time in a single step was $\simeq 6~$minutes, with exception of the last
step for sample GF2. Figure~\ref{figxx} shows the
measured thickness of six different mesoscopic flakes as a function of
the total exposure time to the oxygen plasma. We obtain a similar etching slope
$\simeq 0.7 \pm 0.1~$nm/min for all the samples within the exposed total time and
the used parameters.

When placing the
sample plus substrate into the oxygen plasma chamber, the whole substrate including the
sample as well as the electrodes would be exposed to the plasma. In
order to avoid the destruction of the electrodes through the oxygen etching,
part of the sample and the
electrodes needed to be protected with an insulating material. We deposited
150~nm insulating SiN$_x$-layer on the top of a part of the sample
by means of  plasma enhanced chemical vapor deposition (PECVD).
 SiN$_x$ is insulating and stable
enough to be used as protection for the contacts against oxygen plasma
etching. Even after several etching processes, the SiN$_x$-layer remained
unchanged and its protecting purpose retained. Only a window without the SiN$_x$-layer  was left
(using electron beam lithography with PMMA), in order to etch the
samples between the electrodes afterwards, see Fig.~\ref{tem}.

The temperature dependent resistance and magnetoresistance of the
samples were measured in a commercial $\rm ^4He$ cryostat, within a
temperature range of 2 to 310~K and maximum applied magnetic field
of $\pm~7$~T. Low noise resistance measurements were performed using
an AC bridge (Linear Research LR-700), with the current kept constant
at $5~\upmu$A.

The structural
quality of the mesoscopic graphite samples before and after etching
process was investigated by Raman spectroscopy measurements. For this
purpose, a confocal micro-Raman microscope was used (alpha~$300+$,
WITec) with an incident laser light with $\lambda = 532$ nm and a
maximum power of $35$~mW.

\textcolor{black}{A conventional atomic force microscopy (AFM) device (Veeco, D-3000) with standard AFM tip ($r=30$~nm) was used to measure the surface roughness.
The resolution of the device is limited by the tip radius and it is not possible to resolve point defects.}

\section{Experimental results}

In the following subsections we present and discuss the experimental
results, starting with the Raman results in Section~\ref{sec:raman}.
With this method, we are able to show that the as-received samples are
defect-free. The Raman results obtained after the first thickness reduction show the presence
of defects at the near surface region. This defect contribution, however, does not change with further
thinning and therefore  it is not the reason for the observed changes
in the resistivity presented in  Section~\ref{sec:resis}, and also not in the magnetoresistance in
Section~\ref{sec:mr}.

\subsection{Raman spectroscopy}
\label{sec:raman}

\begin{figure}
\includegraphics[width=\columnwidth]{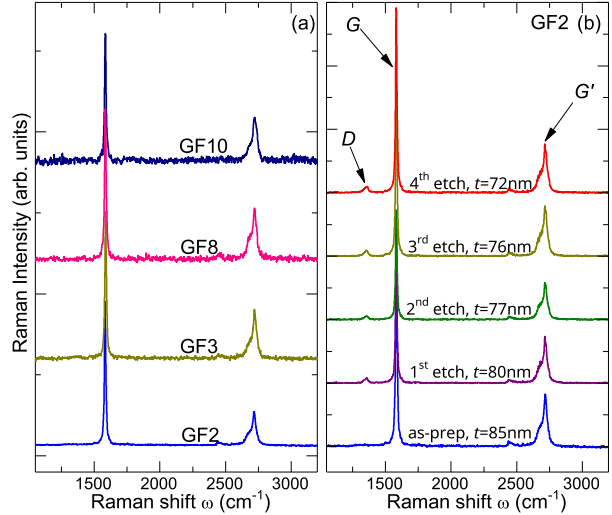}
\caption{\label{fig:Fig3} (a) Raman spectra of some as-received samples before
etching.  (b) The same but for  sample GF2 after several etching steps.}
\end{figure}

Raman spectroscopy (RS) is a powerful method used to investigate the
structure of carbon-based materials~\cite{FERR1,FERR2,DRESS}. With RS
it is possible to investigate samples having micrometer lateral size
and thicknesses from a single layer graphene to bulk
samples. The experimental results of some selected samples are shown
in \figurename~\ref{fig:Fig3}(a).
The effect of the etching process on the Raman spectra is shown
in \figurename~\ref{fig:Fig3}(b) for the case of sample GF2. According
to literature~\cite{FERR1}, the most intense peaks in the Raman
spectra of graphene and graphite are expected at
$\approx 1580~{\rm cm}^{-1}$ (the $G$-peak) and at
$\approx 2700~\rm cm^{-1}$ (the $G'$ peak).  The $G$-peak is due to
the doubly degenerate zone center $E_{2g}$ mode while the $G'$-band is
due to the second order of zone-boundary phonons.  An open question at
the beginning of Raman spectroscopy in graphite was, whether or not it is
possible to detect disorder~\cite{LES,FERR2}. This question was
answered with the observation of the $D$-peak at
$\approx 1350~{\rm cm}^{-1}$, which is related to the disorder present
in the material~\cite{FERR2,FERR4}.  

\begin{figure}
\includegraphics[width=\columnwidth]{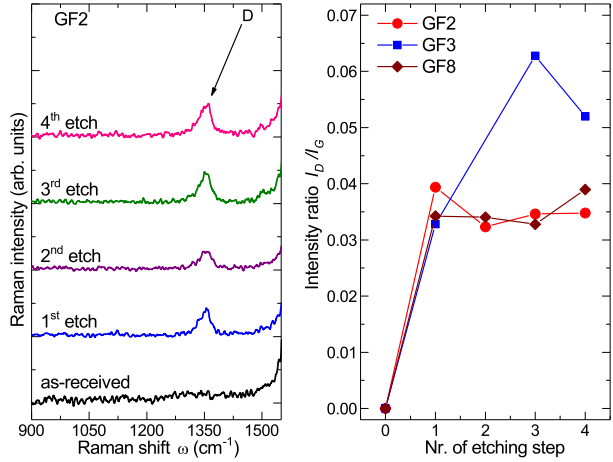}
\caption{\label{fig:Fig4} (a) Raman results of  sample GF2 in the
as-received state and after each
  thickness reduction by oxygen plasma etching. The same data as in \figurename~\ref{fig:Fig3}(b)
  but in an enhanced energy region around the D-peak. (b)
  Intensity ratio $I_D/I_G$ as a function of etching step of three different
  investigated samples.}
\end{figure}

It is important to
 emphasize that the Raman results of the as-received
samples do not show any evidence of disorder, see \figurename~\ref{fig:Fig3}(a) and also \cite{zor17}.
After the first etching, a small peak, the $D$-peak, at
$\omega \approx 1350~\rm cm^{-1}$, appears and remains without changes during all the
 etching steps in each sample, see Figs.~\ref{fig:Fig3}(b) and \ref{fig:Fig4}(a).

In other words, the etching
process results in the appearance of the $D$-peak; the other
peaks do not change after successive oxygen plasma etching within  experimental resolution.
Previous Raman studies on graphene~\cite{CHIL-11}, where the formation of the disorder
peak as a function of oxygen plasma etching  was investigated,
introduced the intensity ratio $I_D/I_G$ as a parameter to quantify
the results.  We have obtained this intensity ratio from our measurements and
we compare them with those from the literature,
see \figurename~\ref{fig:Fig4}(b).

It was shown that after
two-oxygen plasma pulses  $I_D/I_G\approx 0.5$ and the
magnetoresistance shows weak-localization effects~\cite{CHIL-11}, which is typical for
low-disorder single-layer graphene~\cite{MOROZOV-06}.
In our case and after the first etching, the intensity ratio $I_D/I_G$ remains nearly  constant with a
ratio value  at least one order of  magnitude smaller than the ones reported~\cite{CHIL-11}.  
Thus,  from our  Raman results at all etching steps the
observed disorder peak corresponds to a very low density of defects,
 located at the sample
surface. This was already observed at the edges of graphene sheets on the top
of graphite samples~\cite{CANC-04}.
As  we describe above, this disorder at the surface does not change with
further etching procedure and therefore is not
responsible for the occasionally  not systematic changes of the resistance and
magnetoresistance of the investigated samples.

\subsection{Resistance measurements}
\label{sec:resis}
The results of the temperature dependence of the resistance  of
some of the samples, before and after thickness reduction, are presented in
\figurenames~\ref{fig:Fig1} and~\ref{fig:Fig2}. The results of the other samples are
included in the supplementary information.
\begin{figure}
\includegraphics[width=\columnwidth]{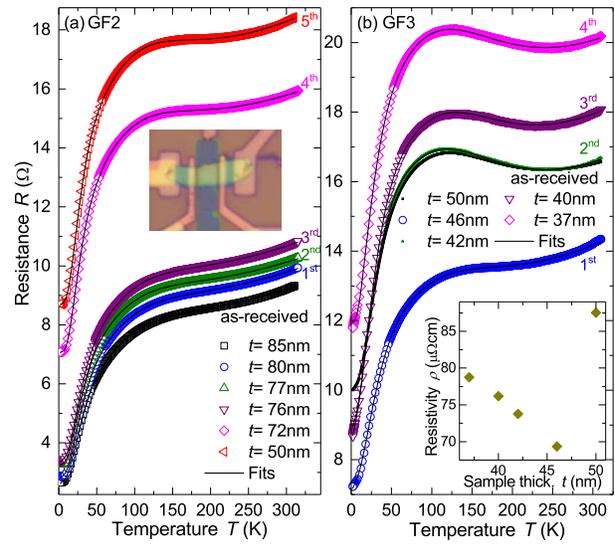}
\caption{\label{fig:Fig1} (a) Resistance vs.~temperature of samples GF2 (a)
  and GF3 (b), before and after etching.  The lines through the points are the fits using
  Eq.~(\ref{eq:Rtot1}). The inset in (a) is an optical image of
  the sample.  (b) In contrast to GF2, sample GF3  is an example for a non monotonous change of the
  resistance after thickness reduction. Note that the as-received curve nearly coincides with the 2$^{\rm nd}$ etching (small symbols). The inset shows the
calculated  resistivity of the same sample at 300~K as a function of the measured thickness. The lines through
the curves are calculated following Eq.~(\ref{eq:Rtot1}).}
\end{figure}
Comparing the results of the untreated samples, different
temperature dependences of the resistance can be observed, although
all samples were obtained from the same initial HOPG bulk material.  Some
samples show a metallic-like behavior over all temperature range, such
as GF2 (\figurename~\ref{fig:Fig1}(a)), others exhibit a maximum and minimum, i.e. like a combination
of metallic and semiconducting contributions, e.g., sample GF3
shown in \figurename~\ref{fig:Fig1}(b) and GF8
in \figurename~\ref{fig:Fig2}(a). In addition, there are some samples with
only a semiconducting-like behavior, as sample GF10
(\figurename~\ref{fig:Fig2}(b)).  Considering that the samples are made of the
same initial material,  that the same experimental procedures were
applied and that the spatial dimensions are also comparable (see
Table~\ref{tab:t1} for detailed information), the overall results already suggest that
the HOPG sample cannot be considered as a homogeneous material.

\begin{table}
\begin{tabularx}{\columnwidth}{p{35pt}XXX}
  \hline\hline
  Sample &  Width ($\upmu$m) &  Length ($\upmu$m) & Thickness (nm)\\
  \hline
  GF2&6.8&9.5&85\\
  GF3&5.3&5&50\\
  GF8&6.7&5.6&72\\
  GF10&8.3&3.8&35\\
  \hline\hline
\end{tabularx}
\caption{Overview of some of the investigated samples and their dimensions,
in the as-received state. The thickness of the samples was measured using
atomic force microscopy, and the other two dimensions correspond to the area
between the electrodes used to measure the electrical resistance.}
\label{tab:t1}
\end{table}

 A dependence of the
estimated resistivity $\rho~(T=300~\rm K)$ on sample thickness is shown in the inset
of \figurename~\ref{fig:Fig1}(b) for sample GF3. The
resistivity decreases after the first etch and then increases reducing the thickness, indicating that the
resistivity of the HOPG graphite bulk cannot be taken as intrinsic.
This kind of behavior of  $R(T)$ and
the thickness dependence of the resistivity was already reported in the
literature with samples of similar quality
but different initial materials and each sample with different current and voltage
electrodes~\cite{oha97,kim05,bar08,barint,zor17}. We note that the estimate of the resistivity 
is done in general using the whole total thickness of the sample. This approach can, upon sample,
be misleading if the electrodes are mainly at the top of the sample. Due to the large anisotropy in the resistance, 
only a smaller part of the total top thickness contributes to the measured voltage.
Decreasing the thickness, 
different internal regions from the rest of the sample start to contribute to the measured voltage leading
to the rather anomalous behavior shown in Figs.~\ref{fig:Fig1} and \ref{fig:Fig2}. 

The results for sample GF2, see \figurename~\ref{fig:Fig1}(a), indicate a
behavior close to that expected for a homogeneous
sample, i.e.~the resistance increases after each thickness
reduction with small changes in the temperature
dependence.   The changes observed for sample GF3 are clearly different from those observed for sample GF2.
The resistance decreases after the
first etching and its temperature dependence changes, see \figurename~\ref{fig:Fig1}(b).
After the second thickness
reduction, the resistance recovers the initial values at
temperatures $T > 20~$K; at low temperatures $R(T)$ behaves different compared to
the initial state.  An interpretation and more detailed discussion of the changes in the
temperature dependence of the samples  will be given in Section~\ref{disc}.

The changes of sample GF8, see \figurename~\ref{fig:Fig2}(a),
are more complicated. After the first
thickness reduction, $R(T)$ changes notably. Moreover, at temperatures
below 150~K the resistance decreases.  A significant change is
observable after the third thickness reduction, where the resistance
is reduced and thus less than after the second etching.  
In the case
of the sample GF10, see \figurename~\ref{fig:Fig2}(b), $R(T)$ behaves
 semiconducting-like above 40~K. After the first etching, $R(T >  50~$K$)$
increases and at $T \lessapprox 30$~K the resistance remains nearly
constant.  

\begin{figure}
\includegraphics[width=\columnwidth]{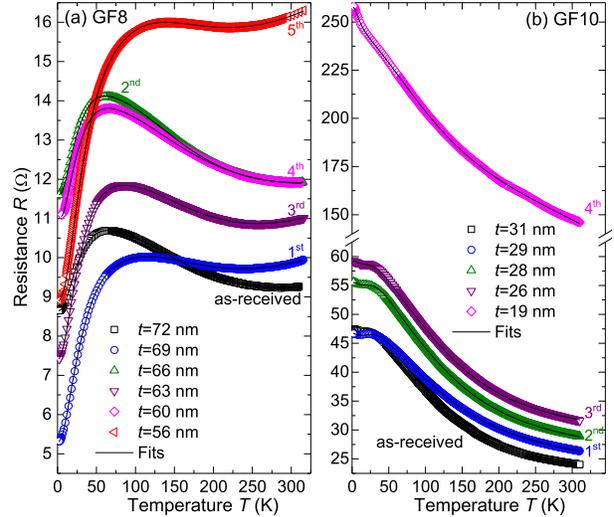}
\caption{\label{fig:Fig2} Resistance vs.~temperature of samples GF8 (a) 
  and GF10 (b), before and after etching.  GF8 shows a varying
  behavior after each etching and GF10 is an example of a
sample with a  semiconductinglike behavior. The lines are results of the
  fits using Eq.~(\ref{eq:Rtot1}).}
\end{figure}

\begin{figure}
\includegraphics[width=\columnwidth]{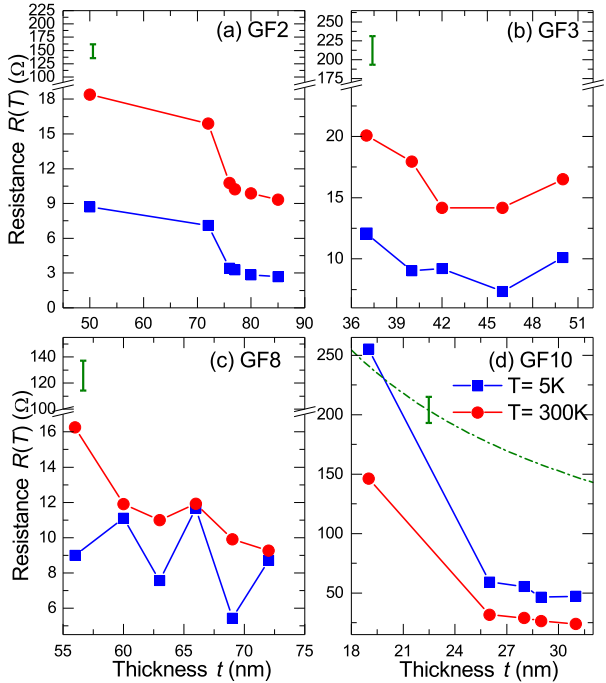}
\caption{\label{fig:Fig6} Resistance measurements of four of the investigated
samples as a
  function of the thickness at 5~K and 300~K. 
  The error bars at the upper left in (a--c) denote the expected
  resistance value for the graphite samples with the smallest achieved thickness 
  with  Bernal phase and without
  interfaces. The dash-dotted  line in (d) represents the resistance calculated assuming a
  constant resistivity of $\rho(5{\rm K}) = 960~\mu\Omega$cm for sample GF10 and its error bar
  applies to the whole curve within the displayed thickness scale.}
\end{figure}

Figures~\ref{fig:Fig6}(a--d) show the change of the resistance
with thickness at  5~K and 300~K. Clearly the resistivity is not
simply inversely proportional to the thickness. 
From previous measurements in
 thin enough samples \cite{bar08,zor17}, where the influence of the interfaces can be
assumed to be minimal, we estimate within $\sim 15\%$ the
intrinsic resistivity $\rho(5{\rm K}) \sim 960~\mu\Omega$cm for 
ideal graphite with Bernal stacking. This is indeed the case for sample GF10, see 
Fig.~\ref{fig:Fig6}(d), which results  we take to estimate the resistivity of nearly ideal graphite with
a Bernal stacking order.  The error bars (in green) in 
Figs.~\ref{fig:Fig6}(a--c) indicate the expected resistance values
at the lowest thickness and at 5~K, were the samples single Bernal phases. 
We note that these estimates of the resistance assumes that the measured voltage comes
from the whole sample thickness. This assumption is expected to be reasonably
correct in thin enough samples where the amount of interfaces and/or 
inhomogeneous regions in parallel do not contribute substantially. 
 In Fig.~\ref{fig:Fig6}(d) we show the estimated resistance vs. thickness
assuming the intrinsic resistivity mentioned above. Interestingly, the
measured resistance at the smallest thickness matches the calculated
one (dash-dotted line in the figure) and the behavior at largest thickness
is nearly  proportional to the expected $1/t$ but shifted downwards by
a constant value of $\sim 100~\Omega$. \textcolor{black}{From all these results we can roughly estimate a penetration depth of $\sim 20$~nm
 for the input current in ideal graphite. However, this estimate should be taken with care because, even without interfaces, lattice defects and grain boundaries in the sample can still increase effectively this penetration.}

\subsection{Surface roughness measured with AFM}

\textcolor{black}{Another possibility to check for the defective state of the surface before and after  oxygen plasma etching is provided by AFM. For sample GF3, AFM was  used also to measure the height and the root mean square (RMS) of the surface roughness of the sample. The results are as follows: In the as-prepared state  the RMS=3.4~nm; after the $1^{\rm st}$ etch RMS=2.5~nm; after the $2^{\rm nd}$ etch RMS=2.1~nm; after the $3^{\rm rd}$ etch RMS=2.2~nm and after the $4^{\rm th}$ etch RMS=2.2~nm. These values indicate that the passive oxygen plasma etching does not increase the surface roughness. Moreover there is no correlation between  the RMS and  the electrical resistance, see Fig.~\ref{fig:Fig6}(b). This is actually expected for a passive plasma etching process, where two oxygen radicals are formed in the plasma, $\rm O^+$ and $\rm O^{2+}$. The resulting compounds, CO and CO$_2$, are then removed by the flow of the process gas and the vacuum pump. There is no acceleration towards the sample, i.e. the reaction happens only at the surface. The removed C-atoms result in dangling bonds at the surface, which can be measured with Raman spectroscopy, see Section 3.1 and Fig.~\ref{fig:Fig4}. Our AFM device does not allow us to measure  point defects with atomic resolution.}

\subsection{Magnetoresistance}
\label{sec:mr}

\begin{figure}
\includegraphics[width=\columnwidth]{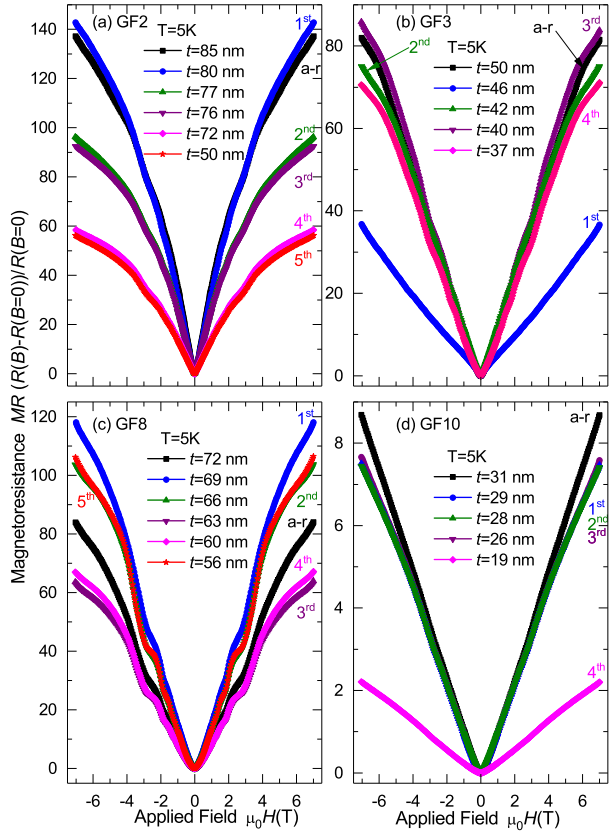}
\caption{\label{fig:Fig5} Magnetoresistance at $T=5$~K of the
  investigated samples presented in Figs.~\ref{fig:Fig1} and  \ref{fig:Fig2}, for
  the as-received (a-r) state and after several etching steps indicated by the
  corresponding numbers.}
\end{figure}

The in-plane MR was measured before and after
reducing the thickness at different constant temperatures and for
fields applied always normal to the graphene layers and
interfaces.  The results at $T=5$~K of some of the investigated
samples are plotted in \figurename~\ref{fig:Fig5}. All samples
regardless of the thickness show a positive MR, before and
after etching. In case of a single graphene layer  the formation
of defects after O-plasma etching~\cite{CHIL-11}  is the
cause for a negative MR measured at low magnetic fields and low
temperatures as a sign of weak localization (WL). A
similar WL contribution and negative MR were observed in few-layers
graphene and interpreted as a consequence of defects present in
the sample~\cite{MOROZOV-06,WANG-12}. To verify  any
contribution of WL in the MR of our samples, we have performed low
field measurements with small field steps in all samples at $T =
5~$K, before and after etching. We did not observe any negative
contribution to the MR. This result indicates that the surface
disorder produced by the etching  does not influence notably the
MR, see Fig.~\ref{fig:Fig5} and corresponding figures in
supplementary information. Considering that this surface
disordered layer should have a much larger resistance than the
graphene and its underlying interface layers, it is expected that its
contribution to the total resistance in parallel  is negligible.

\begin{figure}
\includegraphics[width=\columnwidth]{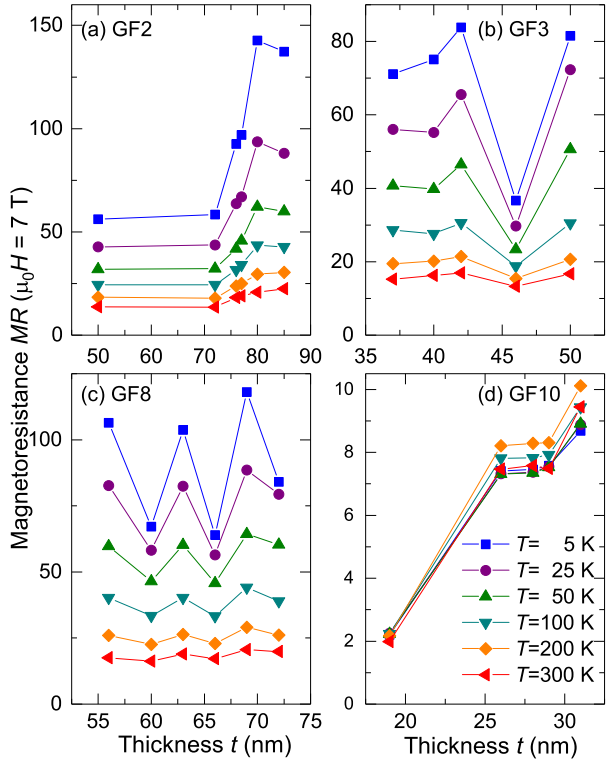}
\caption{\label{fig:Fig7} Magnetoresistance at $\mu_0 H=7$~T of the
investigated samples presented in \figurename~\ref{fig:Fig6} at
different temperatures as a function of the sample thickness.}
\end{figure}

The MR as a function of the sample thickness, i.e. after etching,
follows a general behavior, namely, when the resistance is smaller
the MR is enhanced. The results of the sample GF8 is a good
example to demonstrate this behavior, compare the results in
Fig.~\ref{fig:Fig6}(c) and  Fig.~\ref{fig:Fig7}(c). Early
results~\cite{bar08,barint} showed  that the MR in thick enough
samples is larger compared to thinner samples. Taking into account
the internal microstructure of the samples, from these studies one
concludes that an important part of the measured  MR is directly
related to the influence of a large number of metallic-like
interfaces between crystalline regions.  Similarly, our results
can be understood assuming that at a certain etching process the
defect- or interface-free crystalline layer 
becomes very thin and the contributions of the underlying interfaces increase, see Section~\ref{disc} for details. This suggests that the main contribution to the large
MR comes from internal interfaces or thin defected regions. Further
etching removes the interface completely and the resistance increases
whereas the MR decreases. We note that the MR of graphite is
extremely anisotropic \cite{kempa03}, a fact that speaks for the
contribution of 2D regions.

  In this work we have also
investigated the MR at different temperatures, ranging from 5~K to
300~K, before and after each etching step. In
\figurename~\ref{fig:Fig7} we have plotted the MR at
$\upmu_0H=7$~T vs.~thickness of some of the investigated samples.
In general, the MR decreases, with reducing the sample thickness, as in
samples GF2 and GF10, \figurename~\ref{fig:Fig7} (a) and (d).
However, this behavior is sample dependent because the change of
the MR with thickness depends on whether an interface remains near
the surface region. The MR results of sample GF3, which show a
large decrease in the MR after the first etching and a recovery
afterwards, (\figurename~\ref{fig:Fig7} (b)), and the results of
sample GF8 with its large oscillatory behavior
(\figurename~\ref{fig:Fig7}(c)) are clear results that support the
idea that metallic interfaces formed between crystals strongly
dominate the general electric transport properties. Further
results of other samples can be seen in the supplementary information. In agreement
with the main conclusions obtained in earlier studies
\cite{bar08,arn09,barint}, the whole MR results of the different
microflakes before as well as after etching, indicate the
non-intrinsic origin of the MR.

\textcolor{black}{The large MR in graphite is observed only at fields normal to the graphene layers and
interfaces. At fields applied parallel to them, the MR actually vanishes. The very small, still measurable MR can be quantitatively explained taking into account a small normal field component. This component is due to an intrinsic misalignment between the interfaces and applied field, measured through  
the finite rocking curve width, added to any extra but small experimental misalignment \cite{kempa03}.}

Finally, we would like to note that the observed effects are not
related to a ballistic contribution \cite{esq12}, which would be the case if the electron 
 mean free path is of the order or larger than the sample
size parallel to the graphene and interfaces layers. In the case
of ballistic transport the MR is notably smaller in comparison to
larger samples \cite{gon07,dus11}. If
the ballistic transport overwhelms the diffusive one, the MR {\em
increases} with temperature in contrast to the behavior obtained
in all our samples, see \figurename~\ref{fig:Fig7}. This is
expected because the length and width of our samples are larger
than the mean free path of the conduction electrons ($\ell <
3~\mu$m \cite{esq12}) within the graphene layers as well as at the
interfaces,  in the whole temperature range \cite{esq12}.

\subsubsection{Shubnikov-de Haas oscillations }
\label{sec:sdh}

Shubnikov-de Haas oscillations (SdH) as well as de Haas-van Alphen
oscillations in the transport properties and in the magnetization
as a function of field were considered in the past \cite{KELLY,luky04}
to build a Fermi surface for 3D graphite with different
contributions of electrons and holes of different effective
masses. However, transport measurements of different graphite
samples obtained from Kish graphite \cite{oha00,oha01} with
thickness between 18~nm and 52~nm  indicate that the amplitude of
the SdH oscillations tends to decrease the smaller the thickness
(for a throughout discussion of those results see  \cite{chap7}).
Those results already suggest that there is a non-homogeneous
distribution of patches with a density of carriers large
enough to induce SdH oscillations of similar period in $1/B$.
Taking into account the microstructure of bulk graphite samples,
i.e. the sample-dependent distance between interfaces along the
$c-$axis of the graphite structure, it is appealing to argue that
metalliclike interfaces are the origin for the SdH oscillations.

In a different experiment the SdH oscillations were enhanced
notably by ion irradiation of a 15~nm thick graphite flake, which
hardly showed SdH oscillations before irradiation
\cite{arn09,barnano10}. These early experiments clearly suggest
that the SdH oscillations are not intrinsic of the ideal graphite
structure but they are related to defective regions in the
graphite sample.

%\begin{widetext}

\begin{figure}
\includegraphics[width=\columnwidth]{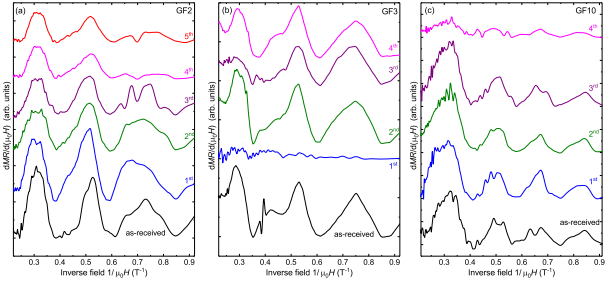}
\caption{\label{GF2} (a) First derivative of the resistance with respect to the 
magnetic field of sample GF2 at different etching steps and
calculated from the curves shown in Fig.~\ref{fig:Fig5}(a). The
$y-$scale is similar for all the curves, which have been shifted
vertically for clarity. (b) Similar to (a)  but for sample
GF3. (c) Similar to (a) but for sample
GF10.}
\end{figure}

%\end{widetext}

For a better characterization of the  SdH oscillations and their
changes with sample thickness, we present the first
derivative of the resistance on field from the experimental curves
shown in Fig.~\ref{fig:Fig5}, i.e. d$R/$d$(\upmu_0 H)$.  We summarize the
results discussing in some detail the results of sample GF2,
GF3 and GF10. 

Figure~\ref{GF2} (a) shows the field dependence of the
first field derivative of the MR obtained for
sample GF2 at 5~K. The SdH oscillations are clearly seen at all
etching steps. It is also observed that the oscillation amplitude
follows the same systematic changes as the MR with thickness, i.e.
the larger the MR, the larger the SdH oscillation amplitude,
compare Fig.~\ref{GF2}(a) with Fig.~\ref{fig:Fig7}(a).

In the case of sample GF3, the MR shows a minimum after the first
etching at a thickness of $t \simeq 46~$nm (see
Fig.~\ref{fig:Fig7}(b)). The clear SdH oscillations observed for
the as-received sample with 50~nm thickness vanish 
completely after the first 
etching step with the resulting thickness of 46 nm, 
see Fig.~\ref{GF2}(b), a remarkable experimental fact that
clearly indicates the non-intrinsic origin of the SdH
oscillations. After further etching the SdH oscillations recover
and maintain a similar amplitude to thickness of 37~nm, similarly
to the MR, see Figs.~\ref{GF2}(b) and \ref{fig:Fig7}(b).

The behavior observed in sample GF10 is similar to the one reported
in previous studies \cite{oha00,oha01}, i.e. the SdH oscillation
amplitude tends to decrease the thinner the graphite sample, see
Fig.~\ref{GF2}(c). The non-systematic change of the MR with
thickness observed in sample GF8, see Fig.~\ref{fig:Fig7}(c), is
also seen in the SdH oscillation amplitude, see supplementary
information, supporting the direct correlation between the value
of the MR and the SdH oscillations amplitude.

\textcolor{black}{The  results obtained from, e.g., sample GF3 (see Fig.~\ref{GF2}(b)) appear to indicate identical carrier densities in the as-received and also after the 2$^{\rm nd}$ etching step. But this is not quite correct. In order to find the $1/B$-frequencies one should use FFT on the whole SdH oscillations curves, because the changes are rather small. For the as-received sample we find 2 main frequencies:  $(0.231 \pm 0.003)$T$^{-1}$ and $(0.120 \pm 0.002)$T$^{-1}$. After the first etching we get $(0.215 \pm 0.005)$T$^{-1}$ and $(0.101 \pm 0.002)$T$^{-1}$ for all other etching steps. Thus, there are differences, though small. Note that  the MR is sensitive to the paths with the largest conductivities. Those can be  located at the regions of relatively large and similar density of states observed in the moir{\'e} patterns formed between twisted layers \cite{kuw90,mil10,flo13,chap7}.}

\section{Discussion}
\label{disc}

In the last years and with the help of scanning transmission electron microscopy
(STEM), it has been shown that HOPG is composed of many crystalline
regions with aligned $c$-axis \cite{bar08} but, either with
different $a$--$b$-axes orientations (i.e.~twisted) with similar
stacking orders on both sides of the interface, or different
stacking orders. Recently, it was reported that in HOPG as well as in natural graphite 
rhombohedral stacking is present with a concentration $\approx 10\% \ldots 20\%$ \cite{pre16}. 
This phase  influences the transport
properties of the whole sample~\cite{zor17} and its interfaces with the Bernal   phase
may show superconductivity, as last studies suggest  \cite{kopbook,mun13,pre16,esq18,chap7,sti18} .
The formation of metallic-like regions within interfaces is not a new concept and
was already observed in many oxide materials, even
superconductivity was found but at very low
temperatures~\cite{REY1,REY2}.

 The crystalline regions inside HOPG bulk have a lateral size in the order of
a few to tenths of micrometer and a thickness varying from ten to hundreds
of nanometers~\cite{gon07,bar08,arn09,esq18}.  We note that the usual thickness
dependence of the resistivity in metallic-like systems, such as
Cu~\cite{KE} or Ag~\cite{TAN}, can be described with the theory of
Fuchs-Sondheimer (FS)~\cite{FUC,SOND}, which considers the
influence of scattering processes at the sample surface. This,
however, is not the case for a material consisting of stacked
layers, such as graphene sheets, where each single sheet is
already conducting~\cite{NOVO,MOROZ}. \textcolor{black}{In other words, the  graphite structure cannot show this surface scattering because the $c-$axis conductivity is several orders of magnitude smaller ($\sigma_c < 10^{-5} \sigma_{a-b}$)  than along the graphene layers. That means that conduction electrons (and holes) do not have an appreciable 
momentum component parallel to the $c-$axis of the graphite structure.  The conduction occurs along the graphene sheets, at which surface scattering does not exist. The absence of weak localization in the MR also indicates that the transport does not occur at the surface. Further, within the Fuchs-Sondheimer model the Matthiessen rule has to be applicable. The electron mean free path in thin graphite flakes was found to be of the order of micrometer and for samples much thinner than $\sim 50~$nm \cite{dus11,esq12} . This implies that the conduction is in plane, i.e. parallel to the surface. If the Fuchs-Sondheimer model would be valid, the mean free path cannot be much larger than the thickness of the samples. Note further that  the resistance of the disordered surface layer should be similar to the one of an amorphous carbon layer. In this case, its contribution to the total resistance (in parallel to the interfaces and graphene layers) is  negligible. This is actually supported by our experimental results where  the  surface disorder measured by Raman and AFM, see Sections 3.1 and 3.3, does not change with further etching procedure and therefore cannot be responsible for the  systematic as well as non systematic changes of the resistance.}

On the basis of early \cite{bar08} and recent \cite{gar12,pre16,chap7}
findings regarding the internal structure of graphite, a simple
parallel resistor model was proposed to describe $R(T)$,
considering the different contributions  from the crystalline
structures and the interfaces~\cite{zor17}. In this work, we  use
this model to describe the results of $R(T)$ of the sample in the
as-received state and after each thickness reduction step, in
order to find some hints on the correlations between the different
contributions and the thickness behavior described above.

\begin{figure}
\includegraphics[width=\columnwidth]{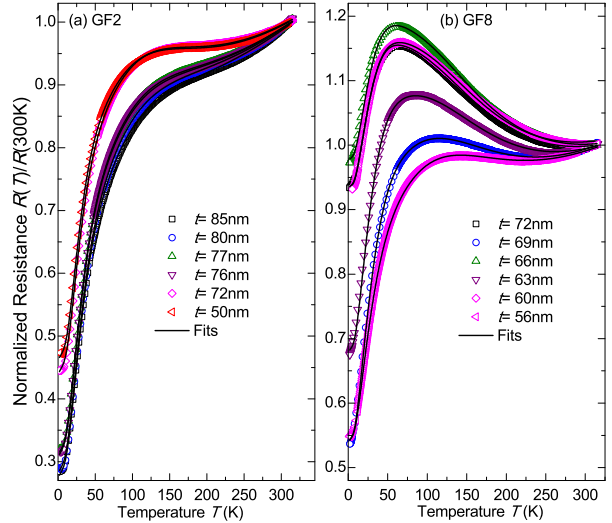}
\caption{\label{fignor} Temperature dependence of the normalized resistances of 
samples GF2 (a) and GF8 (b) at different thickness. The lines are fits to Eq.(\ref{eq:Rtot1}).}
\end{figure}

The model we use assumes that the total measured electrical resistance $R(T)$
consists of three contributions in parallel, one originating from
the interfaces ($R_{\rm i}(T)$) and the other from the crystalline
parts, i.e.~Bernal ($R_{\rm B}(T)$) and rhombohedral ($R_{\rm r}(T)$) stacking orders. 
The total resistance is formulated as
\cite{gar12,zor17}:
\begin{equation}\label{eq:Rtot1}
R(T)^{-1}=R_{\rm i}^{-1}(T)+R_{\rm B}^{-1}(T)+R_{\rm r}^{-1}(T)\,.
\end{equation}
The interface contribution is assumed to behave metalliclike with
a linear and a thermally activated exponential term as discussed in \cite{gar12}:
\begin{equation}\label{eq:Rif}
R_{\rm i}(T)=R_0+ a_i \left [ R_1T+R_2\exp\left( \frac{-E_a}{k_{\rm B}T}
\right) \right ] \,.
\end{equation}
This relation slightly deviates from the originally used in \cite{gar12,zor17}.
The reason is that we would like to have a prefactor $a_i$ that provides the weight  of
this contribution, independent of the constant residual resistance term $R_0$.

The crystalline semiconductinglike contributions for rhombohedral
and Bernal stacking orders are given as usual:
\begin{eqnarray}\label{eq:Rsc}
R_{r}(T)&=&a_1\cdot T^{3/2}\cdot{\rm exp}\left( \frac{+E_{g1}}{2k_{\rm
B}T} \right)\,,\\
R_{B}(T)&=&a_2\cdot T^{3/2}\cdot{\rm exp}\left( \frac{+E_{g2}}{2k_{\rm
B}T} \right)\,.
\end{eqnarray}
\textcolor{black}{Although we do not differentiate between electron and hole carriers to understand  the temperature
dependence of the resistance, the Hall data of different graphite samples clearly indicate the existence of both carriers and should be taken into account \cite{esq14,chap7}}. 
The coefficients $R_0$, $R_1$, $R_2$, $a_i, a_1$ and $a_2$ as well as the
activation energy $E_a$ and the gap energies $E_{g1}$ and $E_{g2}$ are free
parameters we obtain from the fits to the experimental curves. A
detailed discussion of the parameters in terms of their origin and
the fitting procedure can be found in~\cite{zor17}. In the
present work, we use those equations and correlate the fitting
parameters to the changes in the resistance after the thickness
reduction. 

Before we describe the results of the 
fittings to the experimental curves, it is instructive  to plot the results of $R(T)$ of 
samples GF2 and GF8, see Figs.~\ref{fig:Fig1}(a) and \ref{fig:Fig2}(a), in a normalized way, i.e.
$R(T)/R(300)$ vs. $T$, see Fig.~\ref{fignor}. Through this normalization we
can realize better and discuss the changes in the temperature dependence produced by reducing
the thickness of the flakes. The temperature dependence of the resistance of 
sample GF2 remains practically unchanged at the first thicknesses, i.e. from $t = 85$~nm to 76~nm.
\textcolor{black}{However, at the last two
thicknesses 72~nm and 50~nm, see Fig.~\ref{fignor}(a), the sample shows a smaller ratio $R(4)/R(300)$ 
and less metalliclike behavior, i.e. there is a clear  flattening of $R(T)$ between $120~$K$ \lesssim T \lesssim 250~$K. This systematic change of the normalized $R(T)$ with thickness is not observed for sample GF8,  see Fig.~\ref{fignor}(b) and Fig.~\ref{fig:Fig6}(c), although one gets the
impression that both samples show similar trends. The reason for this apparent similarity can be
understood when we realize that upon sample thickness the interfaces,
which provide the metalliclike contribution (see Eq.(\ref{eq:Rif})),
 can be partially removed from the superficial
 region or, after further thickness reduction, a new interface starts contributing.}

\begin{table}
\begin{tabularx}{\columnwidth}{p{58pt}XX}
  \hline\hline
  Parameter &  GF2 & GF8 \\
  \hline
 $E_{g,1}$(meV) & 105 & 113\\
 $E_{g,2}$(meV) & 33 & 41\\
 $E_{a}$(meV) & 8.4 & 6.3\\
 $R_1(\Omega/$K) & $2.3 \times 10^{-3}$ & $1.2 \times 10^{-3}$\\
 $R_2(\Omega)$ & $0.85$ & $0.41$\\
  \hline\hline
\end{tabularx}
\caption{Activation $E_a$ and semiconducting 
gap energies $E_{g1,g2}$ of the selected samples.
$E_{g,1}$ and $E_{g,2}$ are the energy gaps of 
rhombohedral and Bernal stacking orders, see Eqs.~(\ref{eq:Rif},\ref{eq:Rsc}). The 
parameters were
obtained by fitting to Eq.~(\ref{eq:Rtot1}) the experimental  $R(T)$ curves in the
as-received states, before any etching process. The typical
uncertainty we get for the parameters is $\sim \pm 10~\%$. For a brief discussion on the
erros see main text.}
\label{tab:t2}
\end{table}

As an example, we discuss the fittings to the experimental normalized curves of samples GF2 and GF8.
From the fits to the experimental curves of $R(T)$ in the as-received states of the samples, 
continuous lines in Figs.~\ref{fig:Fig1} and \ref{fignor},  we get
the  parameters given in Table~\ref{tab:t2}. Those parameters are kept  fixed for all other
curves obtained at different thicknesses. In other words, the fits to the curves obtained 
after etching are done 
leaving free only the weight prefactors $a_i,a_1,a_2$ and the residual resistance $R_0$.  
We can observe that the fits (continuous lines in Fig.~\ref{fignor}) using
Eq.(\ref{eq:Rtot1}) and with merely  three free parameters, describe very well all the experimental
results. \textcolor{black}{The fixed parameters, i.e. the parameters, which were shared among the etching steps and obtained through the fitting, have usually an error of the order of 1\%.  The strength of our analysis is based on the fact that we can reduce the amount of free parameters through sharing them among different curves taking into account the physics behind.}

\begin{figure}
\includegraphics[width=\columnwidth]{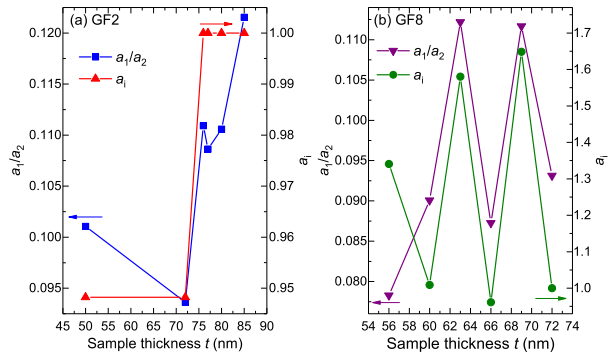}
\caption{\label{figpar} The weight coefficient due to
the interfaces $a_i$ (right $y-$axis) and the ratio of the weight coefficients   of the
rhombohedral and Bernal stacking orders $a_1/a_2$ (left $y-$axis) vs. the
corresponding sample thickness for samples GF2 (a) and GF8 (b), obtained
from the fits to Eq.~(\ref{eq:Rtot1}) setting all parameters of Table~\ref{tab:t2} fixed.}
\end{figure}

\figurename~\ref{figpar} shows the ratio of the weights between different stacking order 
$a_1/a_2$ and the interface weight $a_i$ vs. the thickness of the corresponding samples. 
According to the obtained parameters from the fits the changes in $R(T)/R(300)$ for sample GF2 
can be understood as follows: For the first etching steps down to a thickness of 76~nm, the small
changes are basically due to the small relative decrease of the weight of  the rhombohedral 
phase with respect to the Bernal one. When an interface is removed, i.e. at $t = 72~$nm,  we get 
a clear decrease in $a_i$ with a further decrease in the ratio $a_1/a_2$, 
see Fig.~\ref{figpar}(a). The obtained weight factors for sample GF8, see Fig.~\ref{figpar}(b), indicate
also that a correlation exists between the increase or decrease of the interface contribution with
those from the ratio between the minority rhombohedral stacking order and the majority Bernal; note
that the ratio $a_1/a_2 \sim 0.1$ is in agreement with  XRD studies \cite{pre16}. 
This correlation is somehow expected if
at least some of the interfaces are between the two stacking orders. In other words, removing the region
where an
interface contributes to $R(T)$ implies removing, relatively speaking, 
mainly the minority rhombohedral phase. 

\textcolor{black}{In the case of sample GF10,  the interfaces contribution is diminished 
in comparison to the other samples, as expected due to the small thickness. 
Till the third etch the ratio $a_1/a_2 \simeq 0.08 \pm 0.02$. 
We note that the low temperature behavior was used to fix $R_0$ and $R_1$ ($a_i$ was fixed at 1). The data in the whole temperature
range were then fitted in the usual way, with the energy gaps and the activation energy taken as shared parameters ($E_{g1}=104~$meV, $E_{g2}=41~$meV, $E_a=4~$meV). After the fourth etch, we clear recognize that the behavior of the sample  changed 
dramatically, especially at low temperatures. To get a reasonable good fit of the data for the thinnest sample  we had to leave all parameters  free. 
The  parameters obtained from the fit process suggest that there should be a negligible rhombohedral contribution to the electrical transport, 
together with a clear change in the energy gap of the Bernal phase ($E_{g2} \sim 17~$meV) and 
in the activation energy ($E_a \sim 51~$meV). 
In the supplementary information we compare the results of the fits to the data of sample GF10  with and 
without including the contribution of the rhombohedral stacking at all thickness. 
The obtained differences between the different fit approaches and the experimental data indicate that the contribution of the minority 
rhombohedral stacking is necessary, expect for the sample with $t = 19~$nm, see supplementary information. } 

\section{Conclusion}
We have investigated the electric transport properties of a series
of mesoscopic graphite samples obtained from the same initial bulk
HOPG material. The investigated microflakes are similar in
lateral dimensions but with slightly different thicknesses. By means of
a gentle oxygen plasma etching procedure and the protection with SiN 
the electrodes for the resistance measurements, we were able to investigate
with high precision the influence of thickness
reduction on each sample individually.  Our results show that one can
obtain different temperature dependence upon initial sample and/or
thickness. The reason for the observed changes with thickness is
due to the heterostructure of the graphite bulk samples. This
heterostructure is the reason for the occasionally large changes in the SdH oscillations
amplitude. 

\textcolor{black}{Our results also indicate that those oscillations are not intrinsic of the
graphite ideal structure and
should not be taken as evidence for a 3D Fermi surface. 
By fitting $R(T)$, we confirm the presence of both stacking orders with
semiconducting energy gaps similar to previously reported in the literature \cite{zor17,pam17,hen17}.  
Regarding the reasons for the apparent failure of the well-established models \cite{mcc1,haewal,mcc2,sw58} to predict the obtained energy gaps, we note that they all  
have difficulties to model the van der Waals interaction between graphene layers and they do not include electron-electron or spin-orbit coupling interactions as modern theories do. We further note that these
well-established models do have several free parameters that were obtained from the comparison with experimental transport data without considering the interfaces contribution, see \cite{chap7} and Refs. therein.}

The magnetoresistance  shows also evident changes
induced by the thickness reduction, indicating the significant role of
the interfaces, specially at low temperatures. Raman spectroscopy
confirms that our samples are free from defects in the as-received
state. After thinning using oxygen plasma etching, a
small peak related to defects appears. We correlate this peak to
the defects produced within the first layers on the sample surface
and from the edges, produced during etching. The overall results
indicate that the amount of surface defects
present in the samples after gentle oxygen plasma etching is too small  to
affect the transport and 
they are not the  origin of
observed changes in $R(T)$ or in MR. Our
results provide new convincing
evidence indicating that the transport properties of bulk graphite
are not intrinsic, they depend strongly on the amount of
interfaces present in the material, and the two stable stacking orders of
graphite.
% Finally, XRD measurements in the
%initial bulk HOPG sample reveal the presence of the rhombohedral phase
%in addition to Bernal stacking order.

%\begin{acknowledgments}
Acknowledgments:  We thank Tobias L{\"u}hmann for his help with the Raman spectroscopy
  measurements. Markus Stiller and Jos{\'e} Barzola-Quiquia are
  supported by the DFG collaboration project SFB762. PDE acknowledges the Mincyt of Argentine for the Milstein fellowship and the support of the
Instituto Balseiro in Bariloche, the University of Buenos Aires and the University of Tucum\'an, where part of the manuscript was prepared.
%\end{acknowledgments}

%\bibliography{mgetch,magnetic_carbon_MS}
\bibliographystyle{elsarticle-num}

\newpage

\onecolumn

\large{\bf Supplementary Information of: "Influence of Interfaces on the Transport Properties of  Graphite revealed
by Nanometer Thickness Reduction"}
    \setcounter{section}{0}
\section{Temperature dependence of the resistance  and
  magnetoresistance at $T=5$~K  at different etching steps for samples GF4, GF6 and GF7}
  
In the main article only the results of four samples were shown and discussed, but during this
work we have investigated three more samples, which will be presented
in this supplement.  In~\figurenames~\ref{fig:fig1s}--\ref{fig:fig3s}
the results of samples GF7, GF6, and GF4 are presented. For all
samples  the magnetoresistance MR is plotted in (a), and  the
temperature dependence of the resistance $R(T)$ in (b), at different
etching states, i.e. the same sample after etching and characterized with the
 thickness written in the panels (a). The lines
are the fits using the parallel resistor model, as explained in the main article.  
The insets in the panel (b) show the corresponding MR at 5~K in a smaller field range.

After a few nanometer decrease of the thickness, 
the value and the temperature dependence of the resistance $R(T)$ change. 
But, as we described in the main article, the changes depend very much on
the sample, although all the samples were obtained from the same bulk HOPG
sample. This fact can be easily realized comparing the $R(T)$ curves among the
samples GF7 and GF6 or GF4, see Figs.~\ref{fig:fig1s}(b) and \ref{fig:fig2s}(b) or \ref{fig:fig3s}(b).
The interesting fact of sample GF7 is that the absolute value and the temperature dependence
of $R(T)$ do only slightly  change below 75~K, whereas at higher temperatures the
changes are not monotonous, similar to sample GF8 (see Fig. 6 in the main article).   

\begin{suppfigure}
\includegraphics[width=\columnwidth]{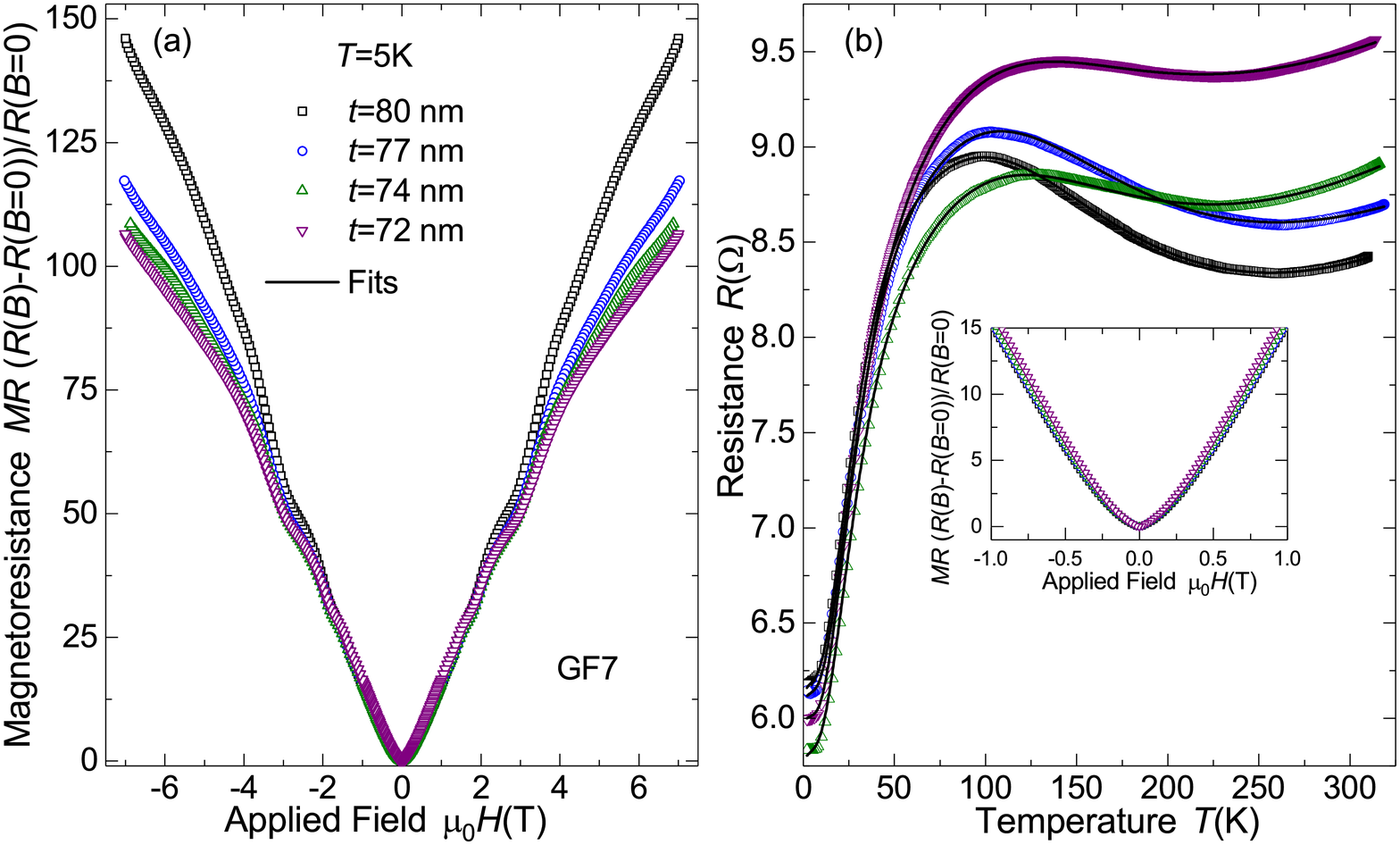}
\caption{\label{fig:fig1s} (a) Magnetoresistance   at $T=5$~K
  and (b) temperature dependence of the resistance at different thicknesses for sample
  GF7. The results of the
  as-prepared sample are always from the sample with the largest thickness. The
  lines shown in (b) are the corresponding fits  using the parallel resistor 
  model as explained in the main article.}
\end{suppfigure}

The MR at 5~K shown in the panels (a) of Figs.~\ref{fig:fig1s},~\ref{fig:fig2s}
and~\ref{fig:fig3s} shows a behavior similar in all samples, i.e.
it decreases the larger the resistance, validating again the conclusion that the MR
and the $R(T)$ are influenced by certain interfaces located near the surface region.
We note that at 5~K the influence of
defects can be observed through the weak localization (WL) effect. This was indeed reported in
graphene and few-layers graphene~\cite{MOROZOV-06,WANG-12}. The plots in the insets 
show no signs of WL   in the MR, neither in the as-prepared state nor after
thickness reduction. These results also indicate that the transport measurements are not
influenced by the surface disordered layer produced by our gentle oxygen plasma treatment.

\begin{suppfigure}
\includegraphics[width=\columnwidth]{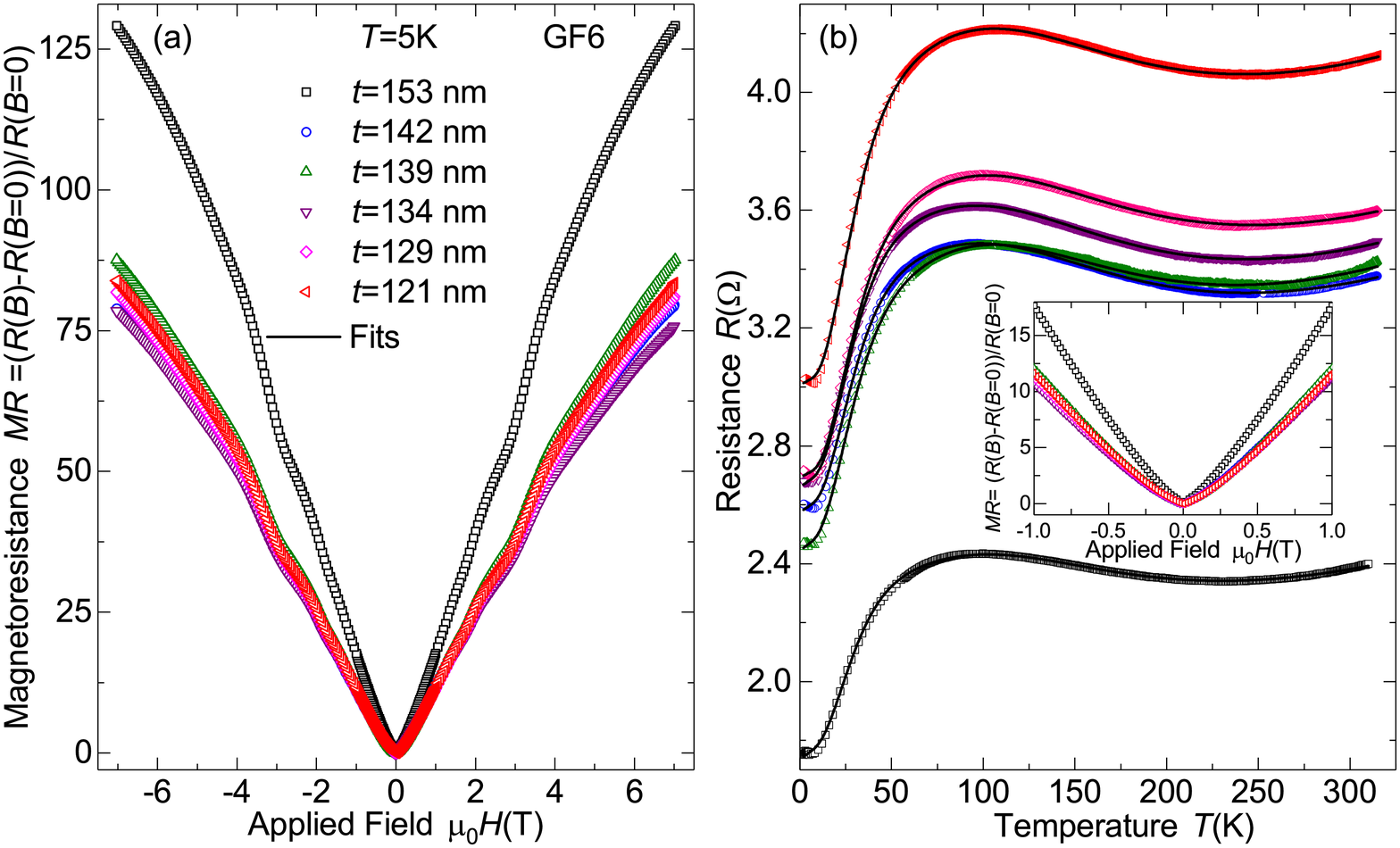}
\caption{\label{fig:fig2s} Similar to Fig.~\ref{fig:fig1s} but for sample GF6.}
\end{suppfigure}

\begin{suppfigure}
\includegraphics[width=\columnwidth]{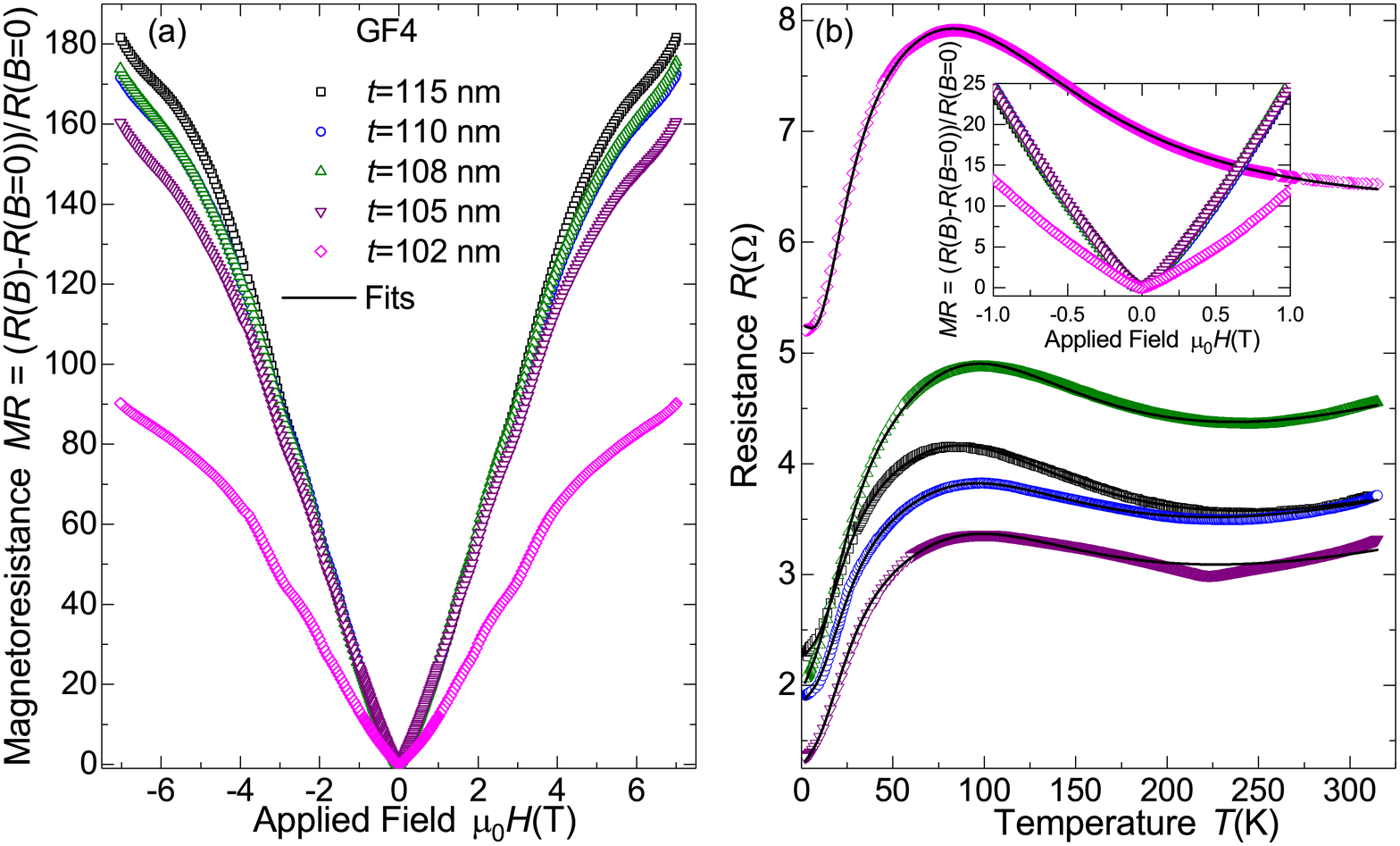}
\caption{\label{fig:fig3s} Similar to Fig.~\ref{fig:fig1s} but for sample GF4.}
\end{suppfigure}

\section{Magnetoresistance at different constant temperatures
  before and after thickness reduction}

The \figurenames~\ref{fig:fig4s} and~\ref{fig:fig5s} show the
magnetoresistance measured at different constant temperatures of
the samples GF3 and GF10 in the as-prepared state (a), and after the
last thickness reduction (b). We show the results of these two samples
as examples of all investigated samples in this work. From the
results we can conclude that in
general the thinning of the samples results in a reduction of the MR.

\begin{suppfigure}
\includegraphics[width=\columnwidth]{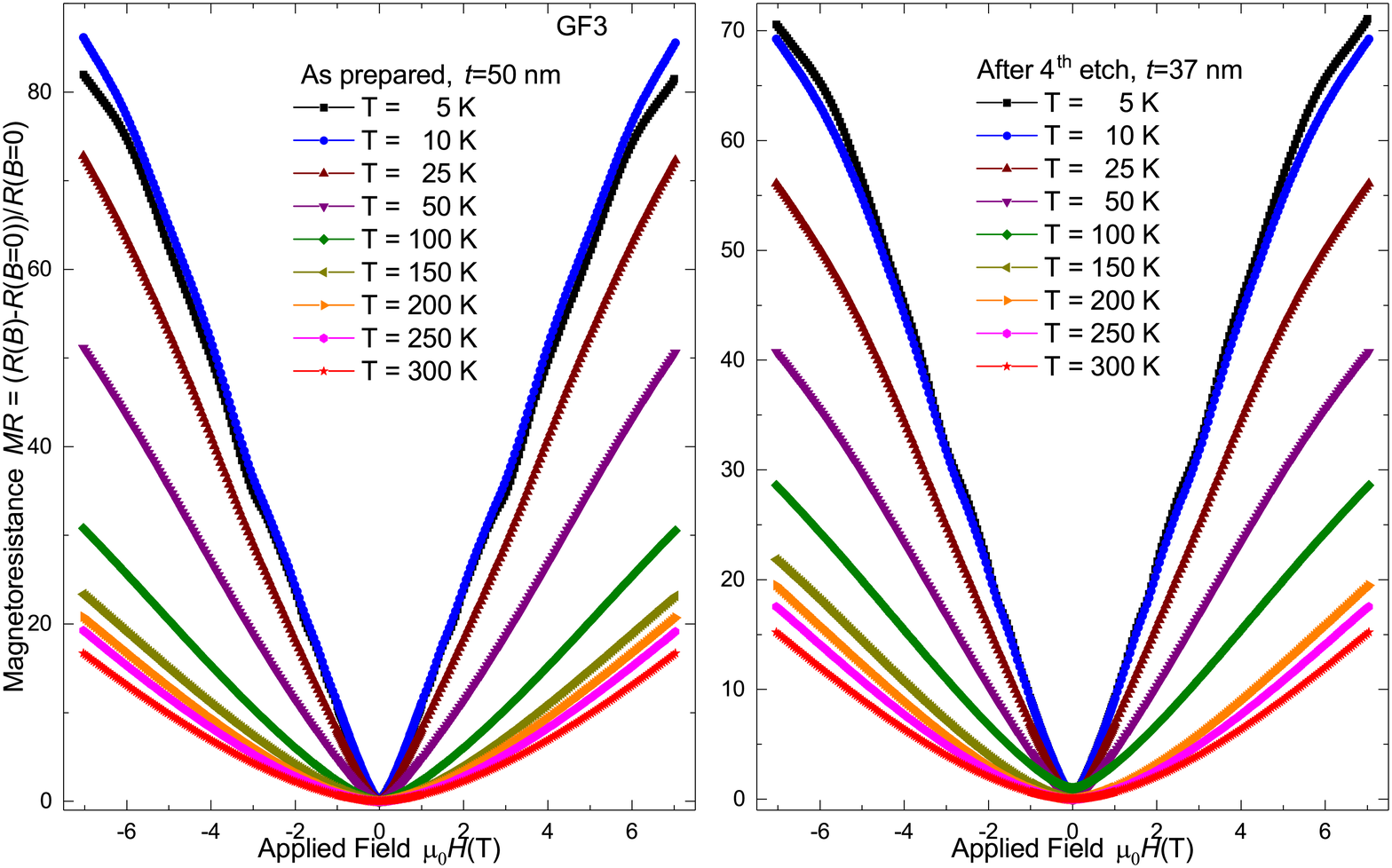}
\caption{\label{fig:fig4s} Magnetoresistance  of  sample GF3
  measured at different constant temperatures in the   as-prepared state (left picture) and   after the fourth etching (right picture).}
\end{suppfigure}

\begin{suppfigure}
\includegraphics[width=\columnwidth]{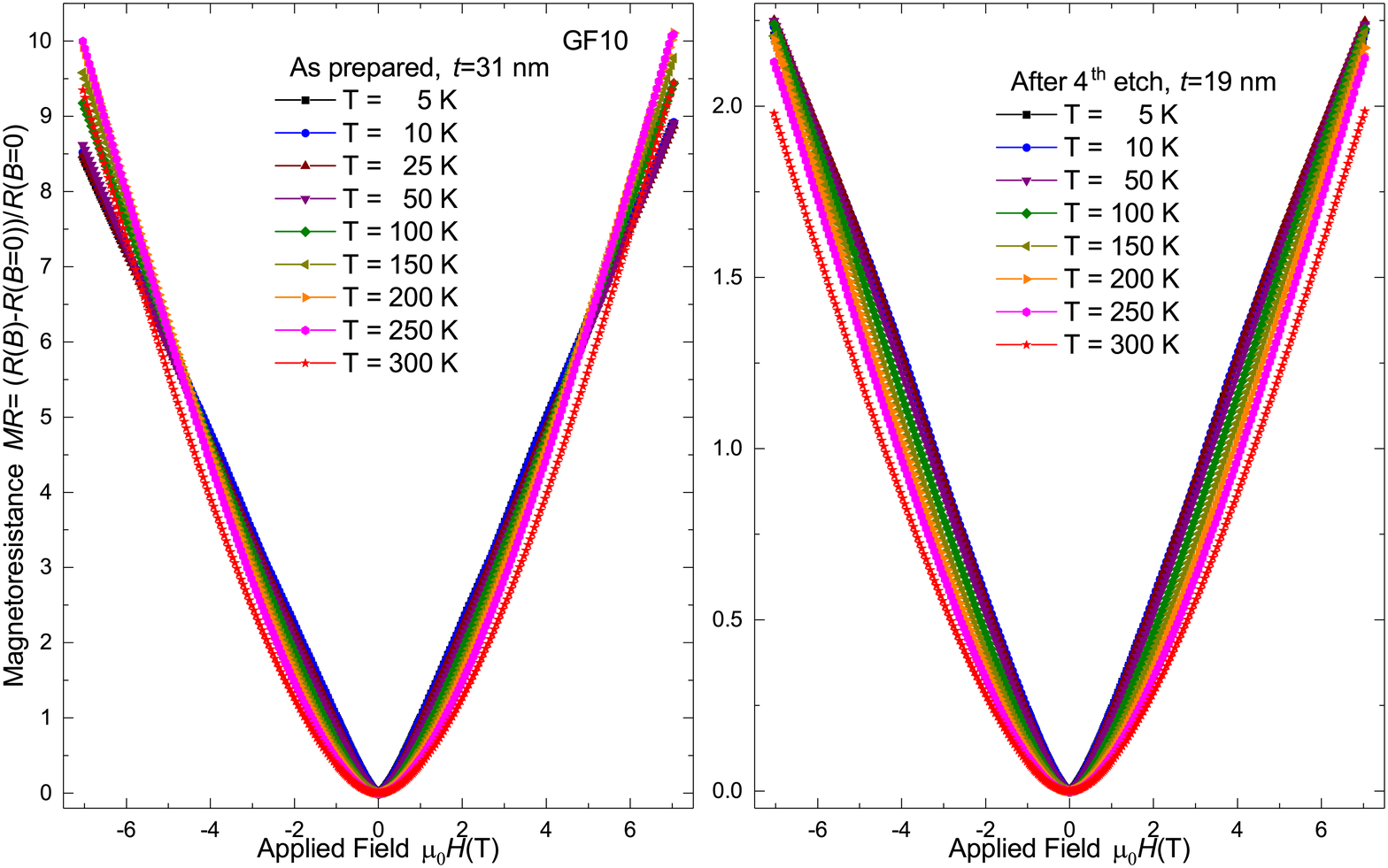}
\caption{\label{fig:fig5s} Similar to Fig.~\ref{fig:fig4s} but for sample GF10. }
\end{suppfigure}

\section{Shubnikov-de Haas oscillations of sample GF8}

Figure~\ref{sdhgf8} shows the field dependence of the
first field derivative of the MR obtained for
sample GF8 at 5~K, see Fig.~8(c) in the main text. 
The SdH oscillations are clearly seen at all etching steps. Their amplitude changes 
with thickness as the  MR at the same temperature, see Fig.~9(c) in the main text. 

\begin{suppfigure}
\includegraphics[width=\columnwidth]{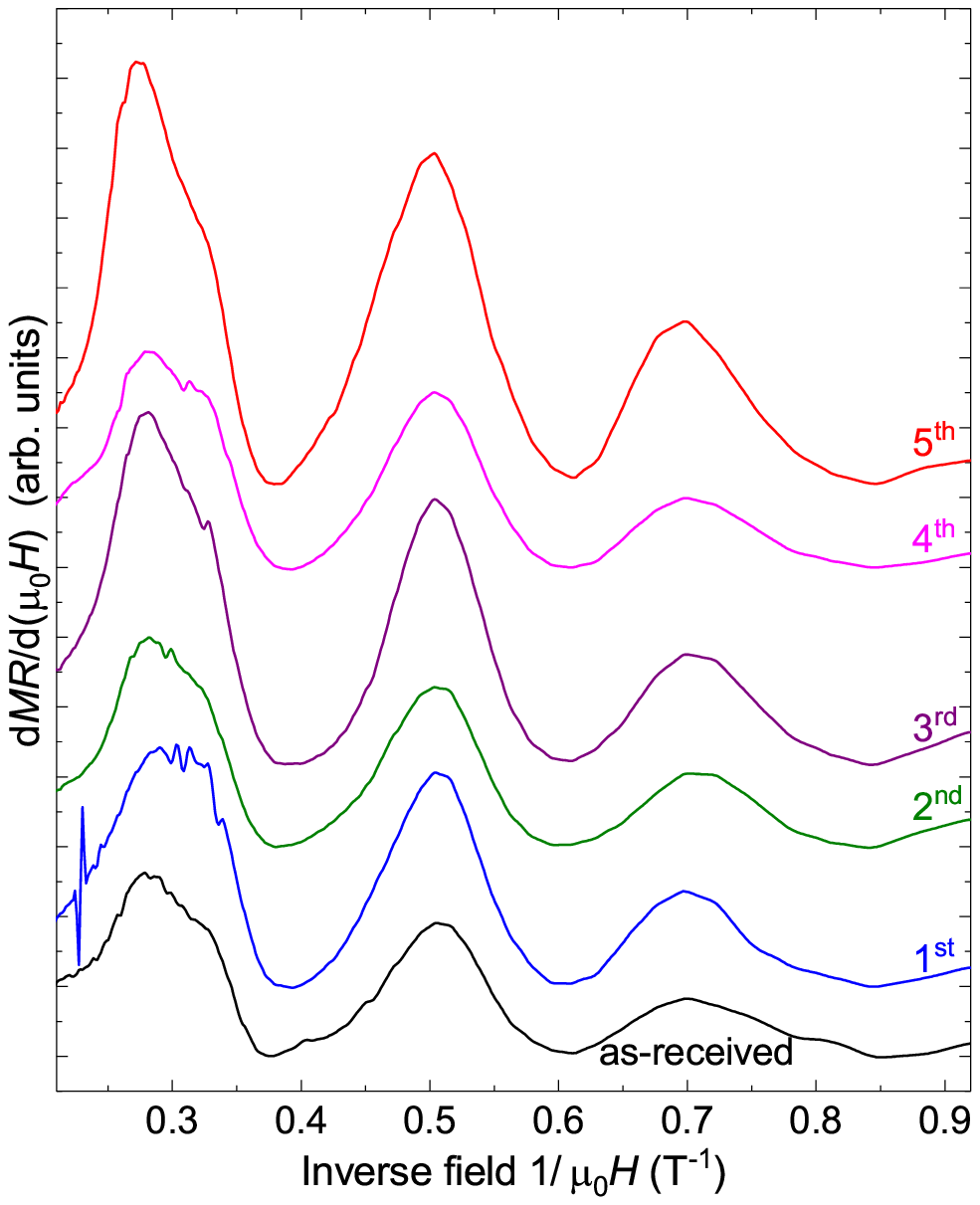}
\caption{\label{sdhgf8}  First derivative of the resistance with respect to the 
magnetic field of sample GF8 at different etching steps and
calculated from the curves shown in Fig.~8(c) in the main text. The
$y-$scale is similar for all the curves, which have been shifted
vertically for clarity.}
\end{suppfigure}

\section{Contribution to the temperature dependence of the minority rhombohedral phase to the resistance of sample GF10}
\begin{suppfigure}
\includegraphics[width=\columnwidth]{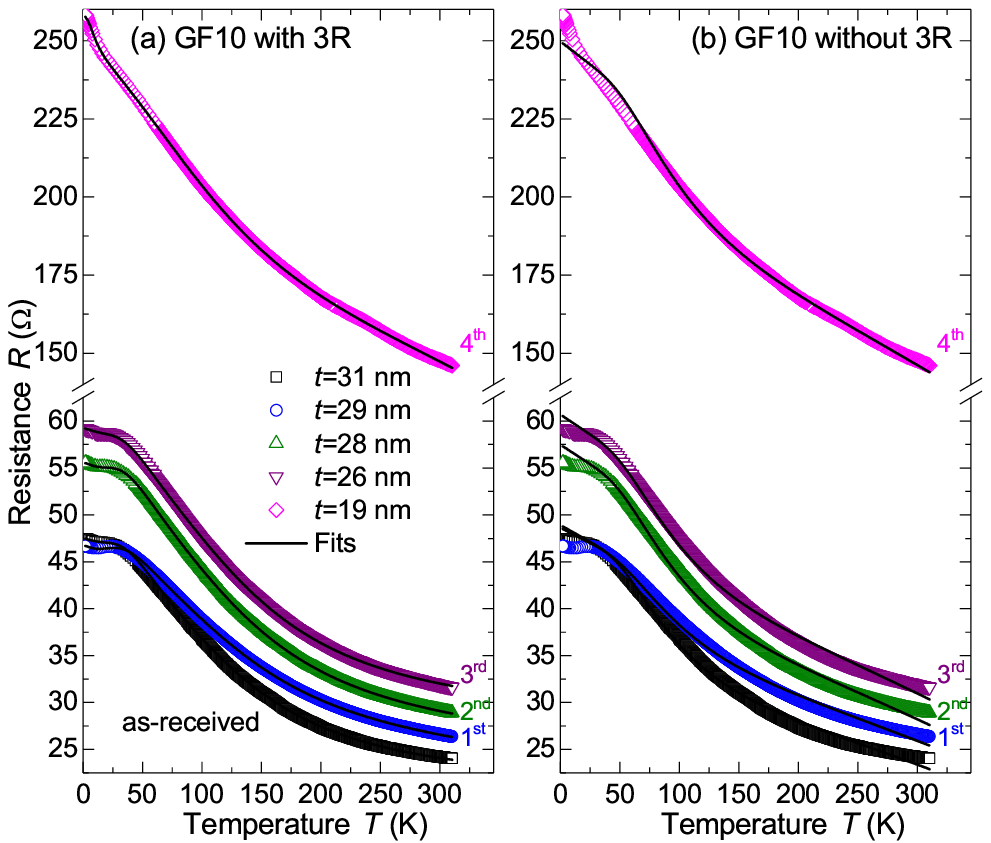}
\caption{\label{comgf10}  Temperature dependence of the resistance of sample GF10 at all thicknesses. The fit curves
in (a) were obtained taking into account the two stacking orders and in (b) without the rhombohedral (3R) stacking.}
\end{suppfigure}

\begin{suppfigure}
\includegraphics[width=\columnwidth]{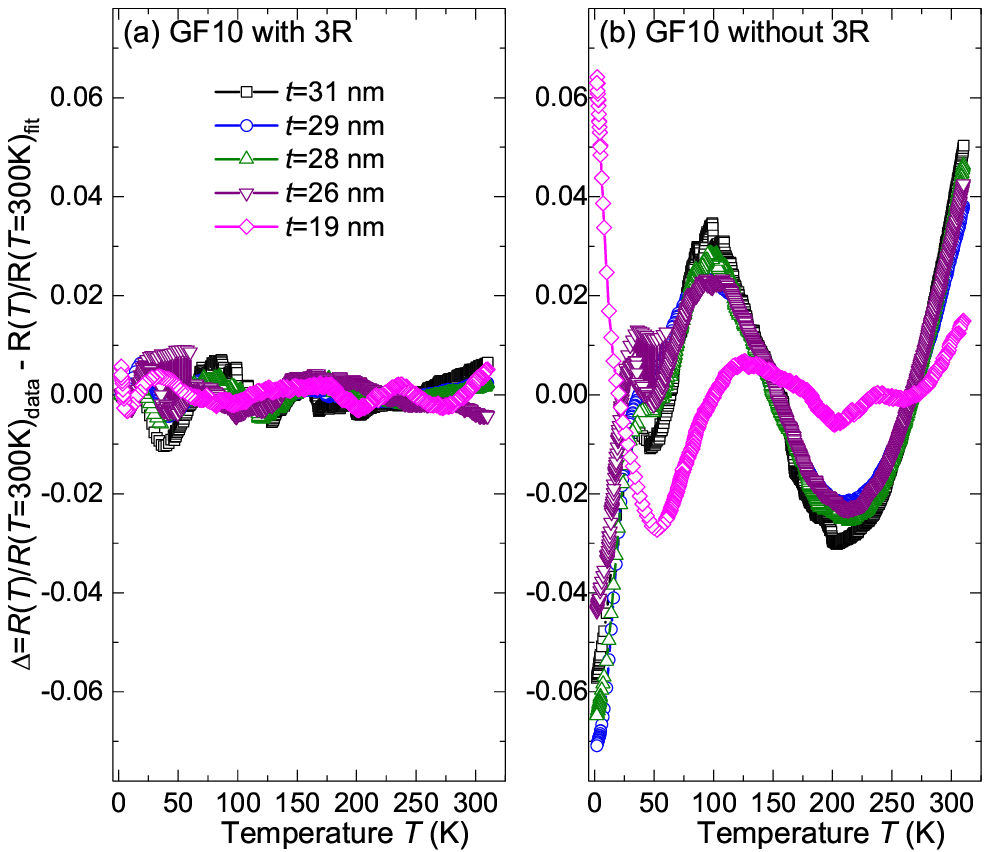}
\caption{\label{resud} Normalized 
residuals $\Updelta=R(T)/R(T=310~\mathrm{K})_{\rm data} - R(T)/R(T=310~\mathrm{K})_{\rm fit}$, including (a) and without (b) the
 rhombohedral stacking (3R). A value of $\Delta \simeq 0.01$ means about 1\% deviation. See similar curves for different samples
 in \cite{zor17-2}.}
\end{suppfigure}

In Fig.~\ref{comgf10}(a) we present the temperature dependence of the resistance of sample GF10 at different thickness 
with the best possible fits including the rhombohedral phase (3R) 
and in (b) without it. As explained in the main text, 
the best fits are found when both stacking orders and an activation term in $R_i$
are taken into account, see Eq.~(1) in the main text, up to the third etch. The linear term is still present with $R_1\approx -0.04 \pm 0.006~\Upomega / \mathrm{K}$ for all samples up to the third etch. To fit the data obtained for the thinnest sample after the fourth etch, 
the rhombohedral phase was not needed. However, a clear change in the energy gap $E_{g2}$ and activation
energy $E_a$ had to be used, see main text.  

We have also tried to fit all the results of sample GF10 without the rhombohedral phase, see Fig.~\ref{comgf10}(b). 
In this case the Bernal energy gap  and activation energy $E_a$ were taken as shared parameters and the other parameters 
were left free. 
Using standard values for  the Bernal energy gap and for the 
activation energy as starting values, the best possible fits were obtained with a reasonable value 
for $E_{g2} \sim 57$~meV (including the  fourth etch), 
but  the contribution of the activation energy term tends to vanish ($R_2 \rightarrow 0$).  Taking the energy gap value $Eg_2$ as free parameter, i.e.~not shared among all curves, did not improve the result. 
As can be seen in Fig.~\ref{comgf10}(b), 
the fits show larger deviations especially at low and high temperatures.

For a better comparison between the fit results, in Fig.~\ref{resud} we have plotted the normalized 
residuals $\Updelta=R(T)/R(T=310~\mathrm{K})_{\rm data} - R(T)/R(T=310~\mathrm{K})_{\rm fit}$, with (a) and without (b) the
 rhombohedral stacking. It is obvious that the deviations are much larger for the case where the rhombohedral stacking was excluded at 
 all thicknesses $t > 19~$nm. We have also tried to manually change the parameters for a better fit, including setting $R_1=0$, yet all trials resulted in even larger residuals $\Updelta$. Note that although in both approaches in (a) and (b) we do not include the
 rhombohedral contribution in the fits at $t = 19~$nm, the difference of the residuals $\Delta$ between the two and for this
 thickness is due to the different values used for the energy gap $E_{g2}$. 
 We may therefore conclude that our fit procedures indicate that the rhombohedral contribution is necessary to fit the data up to the third etch of sample GF10. 
 
%\bibliographystyle{elsarticle-num}

%\bibliography{mgetch,Rhomb,magnetic_carbon_MS}

\end{document}